\documentclass[10pt, twocolumn, aps, prl, reprint, amsmath, amssymb, preprintnumbers, superscriptaddress, longbibliography]{revtex4-2}

\usepackage[utf8]{inputenc}                                             

\usepackage{xcolor}                                                     

\usepackage{amsmath, amssymb}                                           
\usepackage[italicdiff]{physics}                                        

\usepackage{graphicx}                                                   
\usepackage{tikz}                                                       
\usepackage{pgfplots}                                                   
\pgfplotsset{compat=1.14}

\usepackage{hyperref}                                                   

\newcommand*{\lp}{\mathopen{}\left}                                     
\newcommand*{\rp}{\right}                                               
\newcommand*{\hc}{\text{H.c.}}                                          

\renewcommand*{\ket}[1]{\left| #1 \right\rangle}                        
\renewcommand*{\bra}[1]{\left\langle #1 \right|}                        
\renewcommand*{\braket}[2]{\left\langle #1 \middle| #2 \right\rangle}   
\newcommand*{\exv}[1]{\left\langle #1 \right\rangle}                    
\newcommand*{\qexv}[2]{\bra{#1} #2 \ket{#1}}                            
\newcommand*{\norder}[1]{:\hspace{1pt}\mathrel{#1}\hspace{1pt}:}        
\newcommand*{\norderop}[1]{
    \nordbuffopa:\hspace{1pt}\mathrel{#1}\hspace{1pt}:\nordbuffopb}

\newcommand*{\tbuff}{\mathchoice{\quad}{\>}{\>}{\>}}                    
\newcommand*{\kbuff}{\:}                                                
\newcommand*{\eqbuff}{\hspace{1.2em}}                                   
\newcommand*{\nordbuffeq}{\;}                                           
\newcommand*{\nordbuffopa}{\hspace{-0.25em}}                            
\newcommand*{\nordbuffopb}{\hspace{-0.1em}}                             

\newcommand*{\fref}[1]{Fig.~\ref{#1}}                                   

\hypersetup{%
    pdftitle  = {Tunneling times of single photons},%
    pdfauthor = {Jan Gulla, Johannes Skaar},%
    colorlinks,%
    linkcolor={blue!50!black},%
    citecolor={blue!50!black},%
    urlcolor={blue!50!black}%
}

\begin{document}


\title{Tunneling times of single photons}

\author{Jan Gulla}
\affiliation{Department of Technology Systems, University of Oslo, NO-0316 Oslo, Norway}

\author{Johannes Skaar}
\email{johannes.skaar@fys.uio.no}
\affiliation{Department of Physics, University of Oslo, NO-0316 Oslo, Norway}

\date{\today}

\begin{abstract}
Although the group delay of classical pulses through a barrier may suggest superluminality, the information transfer is limited by the precursor which propagates at the vacuum light speed. Single photons, however, have infinite tails, and the question of causality becomes meaningless. We solve this problem by introducing strictly localized states close to single photons, which are examples of optical states produced by on-demand single-photon sources. These states can be arbitrarily close to single photons while demonstrating causality for their leading edge. 
\end{abstract}


\maketitle

\section{Introduction}
For classical pulses, it is well known that the group velocity may be larger than the vacuum light speed $c$ and that the group delay through a barrier may indicate superluminality. Nevertheless, the initial information about the source being switched on is carried by the pulse's precursor, which propagates with speed $c$ because it contains high frequencies not affected by the system. 

When the pulse contains only a single photon, however, things become more complicated. Tunneling times for single particles has been much debated (see, e.g., review articles \cite{hauge1989,winful2006} and references therein), especially since the measurements by Steinberg, Kwiat, and Chiao \cite{steinberg1993}. The issue with single photons is that they have infinite tails falling off almost exponentially or more slowly \cite{bialynicki-birula1998}, which means that they do not have a leading edge. The question of causality for single photons is therefore in principle meaningless. 

On the other hand, there clearly exist on-demand photon sources \cite{wang2019,scheel2009,eisaman2011,senellart2017,sinha2019}. The states produced by such sources must be \emph{strictly localized}, meaning that any local measurement before the leading edge gives the same statistics as for vacuum. Strict localization was introduced in quantum field theory in the 1960s by Knight \cite{knight1961} and Licht \cite{licht1963}. We have recently shown that there exist strictly localized states that approach a single photon when the pulse envelope is slowly varying, and we have estimated the optimal fidelity of near-single-photon states from on-demand sources \cite{gulla2021}.

We take the view that sharp conditions for causality are most easily formulated by having an on-demand photon source and considering local observables as the optical signal propagates through the system. In the Heisenberg picture, it has been shown that photonic pulses from on-demand sources propagate in a causal way \cite{milonni1995}, but it is then difficult to evaluate the photon number of the pulse. In this work, we use the interaction picture and let the optical signal produced by the source be a strictly localized near-single-photon state with an explicit representation from \cite{gulla2021}. By considering how this state propagates through an optical system, we can establish causality and tunneling times for states arbitrarily close to single photons. 

The analysis is done in the framework of quantum optics, i.e., with the full, quantized electromagnetic field and with the interacting matter treated as macroscopic averages. There have also been other methods used to understand photon causality, tunneling, and localization, particularly within quantum electrodynamics \cite{milonni1994,milonni1995,ferretti1968,keller1999,bialynicki-birula1998,saari2005,knight1961,licht1963}.

\begin{figure}[tb]
    \includegraphics{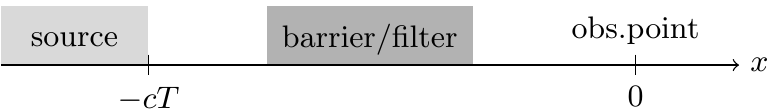}
    \caption{\label{fig:setup}A source is located in the region $x < -cT$, and we consider field observables at the observation point $x = 0$ on the other side of a barrier or an optical filter. The source is turned on at $t = -T$. By causality, the measurement statistics for $t < 0$ must be equal to those for vacuum.}
\end{figure}

The setup in our analysis (see \fref{fig:setup}) consists of a plane-wave (1D) source located in the region $x < -cT$, where $T > 0$ is a constant. For time $t < -T$, the electromagnetic field is assumed to be in the vacuum state $\ket{0}$. The source is turned on at $t = -T$ and off at $t = -T/2$, producing a state $\ket{\eta}$. We consider field observables at the observation point $x = 0$ on the other side of a barrier or an optical filter. Since we assume the source is operated on demand, the state $\ket{\eta}$ must be strictly localized to $x \leq ct$, and at the observation point it is therefore localized in time to $t \geq 0$. 

\section{Strict vs. weak localization}
\begin{figure}[tb]
    \includegraphics{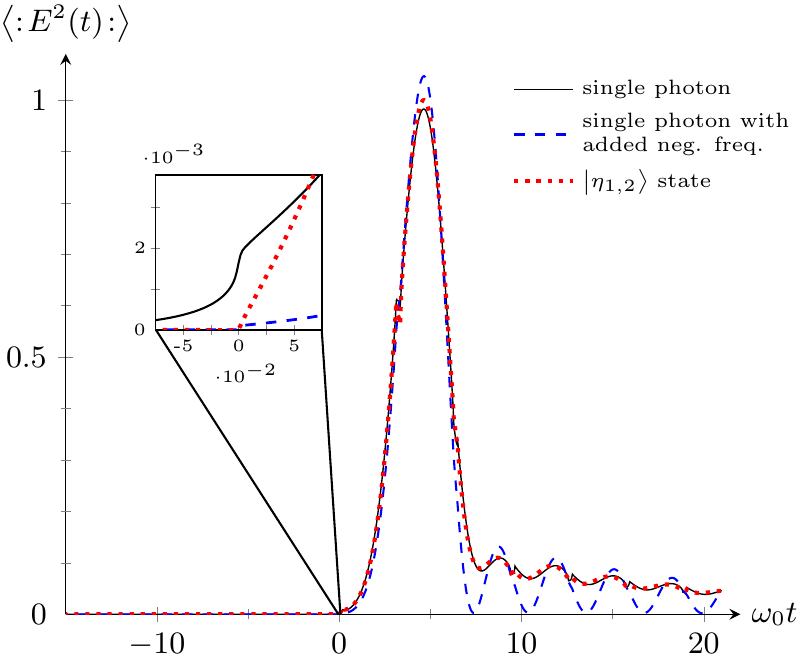}
    \caption{\label{fig:comparison}Energy density (arb. units) as a function of time at the output of a Fabry-Perot interferometer when the incident state is produced by a very fast single-photon source. The definitions of the various quantities are made precise later in the manuscript. The curves show the calculation for 3 different representations of the incident pulse: The black solid line represents an incident pulse containing exactly one photon, where the energy density is given by the conventional spectral integral over only positive frequencies. The dashed blue line represents an incident single photon where the energy integral is extended to negative frequencies as well. The dotted red line is the expected energy density of an incident near-single-photon state $\ket{\eta_{1,2}}$ presented in this work. The time $t = 0$ corresponds to when an input signal propagating freely at speed $c$ reaches the output.}
\end{figure}

To clarify the goal of this work, it is helpful to have an early, \emph{qualitative} look at one of the results, shown in \fref{fig:comparison}. The definitions and precise methods for calculating the various quantities will be presented later in the manuscript. The figure shows numerical plots of the energy density as a function of time at the observation point when the optical filter is a Fabry-Perot interferometer. The energy is calculated for 3 different representations of the state produced by an ultrafast single-photon source. 

The conventional representation of the state is simply a pure single-photon pulse. For a 1D, single-polarization source located to the left of the observation point, we can assume that the pulse contains only one mode (wave vector) per frequency \footnote{For calculating normal-ordered expectation values, we may omit modes in the field expression for which the state contains vacuum. Choosing to omit leftward-moving modes is in principle no different than omitting the other 3D directions when considering 1D propagation; it is anyway just a choice since we want to find an example of a strictly localized state for which the other modes are vacuum. There could very well be other solutions of states strictly localized to $t \geq 0$ that contain both left- and right-moving components (or other 3D modes), but we show that solutions exist with purely rightward-moving modes.}. By noting that a quantum particle's frequency (energy) is always positive, it follows that the single-photon energy density as a function of time is expressed by a function with only positive frequencies, meaning that it must decay slower than exponentially by the Paley-Wiener criterion \cite{paley1934,bialynicki-birula1998}. Therefore, as illustrated by the solid line in \fref{fig:comparison}, the single-photon energy density takes nonzero values for $t < 0$, giving a violation of causal signal propagation that is, however slight, in principle always present.

This does not contradict the well-studied topic of highly localized optical pulses. There is a large amount of known solutions of electromagnetic waves with strong localization properties, also in the case of single photons \cite{newton1949,jauch1967,amrein1969,bialynicki-birula1998,saari2005,adlard1997,ciattoni2007,saari1997}. It is even known that for a 3D single photon, one may obtain a localization that is stronger than exponential in one direction \cite{saari2005}, but at the expense of weaker localization along another directions. However, these are all examples of \emph{weakly localized} solutions, where the energy density falls off rapidly but never reaching exactly zero. In particular, for a 1D, single-polarization single photon, the limitations of the Paley-Wiener criterion cannot be avoided. 

One method suggested to remedy this problem has been to extend the spectral integral from the single-photon expression to include negative frequencies as well (see, e.g., \cite{loudon2001}, Ch.~6.2). By Fourier transforming the single-photon spectrum, truncating for negative $t$, and transforming back, we obtain a spectrum containing all frequencies. The energy density as a function of time for such a state is then identically 0 for $t < 0$ (dashed line in \fref{fig:comparison}), consistent with causality. However, even though this method is a good approximation when the pulses are slowly varying, it is not really physical; negative frequencies of a single quantum particle are not real. 

Thus, there is an issue with how to describe single-photon states produced by on-demand sources. First, it is clear that such states in fact cannot only contain exactly one photon. Nevertheless, as evidenced by reported experimental realizations (e.g., \cite{wang2019}), we expect that on-demand sources can produce states \emph{close} to single photons. As such, we consider states close to exact single photons as representing the true emitted signal by single-photon sources. Second, if the goal is instead to simply find an expression for the energy density of the output state that is ``good enough'', there are no candidates that convincingly give the correct behavior around $t = 0$, which is needed to analyze questions of causality.

We address both of these problems in this work. Based on our earlier result \cite{gulla2021}, we describe the construction of a state $\ket{\eta_{1,2}}$ that is very close to a single photon as measured by the fidelity. At the same time, we show that this state can be produced by an on-demand source since it is indistinguishable from vacuum for all $t < 0$. We then find an exact expression for the time dependence of the expected energy density of this state, as shown with the dotted line in \fref{fig:comparison}. This curve closely matches the expected energy density for a pure single photon for large $t > 0$ but deviates around $t = 0$, instead exhibiting the appropriate, causal behavior. We emphasize that the state $\ket{\eta_{1,2}}$ is not constructed to represent some novel photonic source but rather is suggested to be a better representation of the true state produced by any on-demand single-photon source. Once a proper representation of the state is found, we proceed with the main goal of rigorously establishing strict causality for single-photon tunneling.

\section{Local observables}
The key requirement for any strictly localized state $\ket{\eta}$ is that it must be indistinguishable from vacuum for $t < 0$, meaning that any observable \emph{local} to $t < 0$ must give the same expectation value for $\ket{\eta}$ as for vacuum. Local observables were categorized by Knight \cite{knight1961} as those obtained by sums and products of the field operator evaluated in the measurement region (as well as constants). For example, defining $E(t)$ as the interaction-picture electric field operator at the observation point $x = 0$, we may consider the local observable $E^2(t)$ or $\norder{E^2(t)}$, where $\norder{}$ denotes normal order. The latter is proportional to the energy density (minus the vacuum contribution, which is a constant) and is local to the time $t$ of the measurement. This observable was also used by Bialynicki-Birula in his discussion of near-exponential photon localization \cite{bialynicki-birula1998}. A necessary requirement for $\ket{\eta}$ is thus
\begin{equation}\label{expval_vacuum}
    \qexv{\eta}{\norder{E^2(t)}} = \qexv{0}{\norder{E^2(t)}} = 0, \tbuff t < 0. 
\end{equation}

It is important to note that in principle, the traditional intensity operator $E^-(t) E^+(t)$ in quantum optics (\cite{loudon2001}, Ch.~4.11) is \emph{not} a local observable \cite{tatarskii1990,milonni1995}, since $E^\pm(t)$ are not locally related to the electric field. Here, $E^\pm(t)$ are the positive- and negative-frequency components of $E(t)$. Thus, the intensity operator can give tails even for strictly localized states. This artifact is a consequence of applying the rotating-wave approximation in Glauber's detection theory \cite{milonni1995}, which is not warranted when discussing causality and precursors \footnote{The inapplicability of the intensity operator for rapidly varying envelopes is not so different from the situation in classical electrodynamics. For example, writing the electric field as $E(t) = \Re{E_0(t) e^{-i\omega_0 t}}$ with carrier $\omega_0$ and envelope $E_0(t)$, the electric energy density is $\frac{1}{2} \epsilon_0 E^2(t) = \frac{1}{8}\epsilon_0 \lp[E_0(t)^2 e^{-2i\omega_0 t} + E_0^*(t)^2e^{2i\omega_0 t} + 2E_0^*(t) E_0(t) \rp]$, where $\epsilon_0$ is the permittivity. In the quasi-monochromatic case where the envelope $E_0(t)$ varies slowly, we can average over an optical period such that the first two terms are washed out. This leads to a result proportional to $E_0^*(t) E_0(t)$. However, when the envelope varies rapidly, we must retain the first two terms, keeping the original expression $\frac{1}{2}\epsilon_0 E^2(t)$.}. For another discussion of local observables in quantum optics, see the Supplemental Material in \cite{gulla2021}. 

These considerations must be taken into account even if they happen to be practically out of reach with today's technology. The regime in which Glauber's intensity operator $E^-(t) E^+(t)$ gives an appreciable error is of course rather extreme: the optical pulse length must be on the order of a single cycle. Detecting the rapid changes in such a wide-spectrum signal would require a photodetector with a similarly broad bandwidth, something which seems not yet attainable with current detection methods. However, these are purely technological limitations that will surely improve over time, and proposals for single-photon sources with a pulse length as short as a single cycle are appearing \cite{su2016}. For a sufficiently fast pulse and detector, the measurement outcome will no longer be accurately described by Glauber's intensity operator, and it must instead be characterized by an appropriate local observable.

\section{Pulse modes}
It turns out to be useful to introduce a complete, orthonormal set of pulse modes $\lp\{\xi_n(\omega)\rp\}_n$, supported for frequencies $\omega \geq 0$. We can then define corresponding pulse-mode ladder operators
\begin{equation}\label{eq:a_n}
    a_n^\dagger = \int_0^\infty \dd\omega \xi_n(\omega) a^\dagger(\omega),
\end{equation}
where $a^\dagger(\omega)$ is the usual continuous-mode creation operator (\cite{loudon2001}, Ch.~6.2). The pulse-mode ladder operators satisfy $\comm{a_n}{a_m^\dagger} = \delta_{nm}$. 

We consider a plane-wave source of a single polarization which is located to the left of the observation point. This means that we can assume that the state produced by the source only contains rightward-moving wave vectors in 1D \cite{Note1}, i.e., one wave vector per frequency $\omega$. Thus, we can write the interaction-picture electric field operator at $x=0$ as an integral over the positive frequencies (see also \cite{loudon2001}, Ch.~6.2),
\begin{equation}\label{eq:E_pos}
    E(t) = \int_0^\infty \dd\omega \mathcal{E}(\omega) a(\omega) e^{-i \omega t} + \hc,
\end{equation}
where we for later convenience write $\mathcal{E}(\omega) = K \sqrt{-i\omega}$ with a constant $K > 0$, absorbing the additional phase factor into $a(\omega)$. Using \eqref{eq:a_n}, we can rewrite the electric field operator in the pulse-mode basis,
\begin{equation}\label{eq:E_pos_n}
    E(t) = \sum_n E_n(t) a_n + \hc,
\end{equation}
with associated functions
\begin{equation}\label{eq:E_n_def}
    E_n(t) = \int_0^\infty \dd\omega \mathcal{E}(\omega) \xi_n(\omega) e^{-i\omega t}.
\end{equation}
Note that according to the Paley-Wiener criterion \cite{paley1934}, the functions $E_n(t)$ must be nonzero (almost) everywhere since they contain only positive frequencies. In particular, they have infinite tails for $t < 0$.

\section{Strictly localized state \texorpdfstring{\\}{}near single photon}

Ref. \cite{gulla2021} gives the following algorithm for constructing a state $\ket{\eta_{1,2}}$ that is strictly localized to $t \geq 0$ while also being close to a single photon:
\begin{enumerate}
    \item Pick a complex-valued, square-integrable function $g(t)$ with $g(t) = 0$ for $t < 0$, and calculate its Fourier transform $G(\omega)$. We will refer to $g(t)$ as the seed function for the state.

    \item\label{itm:G_tilde} Modify $G(\omega) \mapsto \widetilde{G}(\omega)$ as follows:
    \begin{equation}\label{eq:G_tilde}
        \widetilde{G}(\omega) = G(\omega) - \beta G^*(-\omega),
    \end{equation}
    where 
    \begin{align}
        \beta &= \frac{1}{2I^*} \lp(1 - \sqrt{1 - 4 \abs{I}^2}\rp), \\
        I &= \frac{\int_0^\infty \dd\omega G(\omega)G(-\omega)}{\int_{-\infty}^\infty \dd\omega \abs{G(\omega)}^2}.
    \end{align}
    
    \item\label{itm:G_xi} Normalize $\widetilde{G}(\omega)$ such that $\int_0^\infty \dd\omega \abs*{\widetilde{G}(\omega)}^2 = 1$. Identify two pulse-mode spectra $\xi_1(\omega)$ and $\xi_2(\omega)$ using
    \begin{subequations}\label{eq:xi_def}
    \begin{alignat}{3}
        \xi_1(\omega) &= \widetilde{G}(\omega), \tbuff &&\omega > 0, \\
        \xi_2(\omega) &= \sqrt{\frac{1 - \eta}{\eta}} \widetilde{G}(-\omega), \tbuff &&\omega > 0.
    \end{alignat}
    \end{subequations}
    The constant $\eta > 0$ is picked such that $\xi_2(\omega)$ gets normalized. The two pulse modes are orthogonal because of step \ref{itm:G_tilde} and can therefore be chosen as two modes in the basis $\lp\{\xi_n(\omega)\rp\}_n$. 
    
    \item Define operators
    \begin{align}
        A_1^\dagger &= a_1^\dagger \frac{1}{\sqrt{a_1 a_1^\dagger}} = \sum_n \ket{{n + 1}_1}\bra{n_1}, \label{eq:A_1_def} \\
        S &= e^{\gamma a_1 a_2 - \gamma a_1^\dagger a_2^\dagger}, \label{eq:S_def}
    \end{align}
    where $\ket{n_k} = {a_k^\dagger}^n \ket{0_k} / \sqrt{n!}$, $\ket{0_k}$ is the vacuum state of pulse mode $k$, and $\tanh\gamma = \sqrt{\eta/(1 - \eta)}$. Our strictly localized state is
    \begin{equation}\label{eq:eta_state_def}
        \ket{\eta_{1,2}} \equiv S^\dagger A_1^\dagger S \ket{0}.
    \end{equation}
\end{enumerate}

By expanding the exponentials of $S$ and $S^\dagger$, we find that the state $\ket{\eta_{1,2}}$ is of the form
\begin{equation}\label{eq:eta_state_expansion}
    \ket{\eta_{1,2}} = c_1 \ket{1_1 \kbuff 0_2} + c_2 \ket{2_1 \kbuff 1_2} + c_3 \ket{3_1 \kbuff 2_2} + \dotsb,
\end{equation}
for coefficients $c_1, c_2, \dotsc$ Its fidelity with the single-photon state $\ket{1_1 \kbuff 0_2}$ is given by the first coefficient,
\begin{equation}\label{eq:eta_state_fidelity}
\begin{aligned}
    F   &\equiv \abs{\braket{1_1 \kbuff 0_2}{\eta_{1,2}}} = \abs{c_1} \\
        &= 1 - \lp( \frac{3}{2} - \sqrt{2} \rp) \eta + \order{\eta^2} \approx 1 - 0.09\eta,
\end{aligned}
\end{equation}
where the approximation is valid for small $\eta$. 

We see that the parameter $\eta$ is determined by the choice of seed function $g(t)$ of the localized state and that it is given by the fraction of the squared norm of $\widetilde{G}(\omega)$ for negative frequencies,
\begin{equation}\label{eq:eta_def}
    \eta = \frac{\int_{-\infty}^0 \dd\omega \abs*{\widetilde{G}(\omega)}^2}{\int_{-\infty}^\infty \dd\omega \abs*{\widetilde{G}(\omega)}^2}.
\end{equation}
The value of $\eta$ quantifies the state's similarity with a single photon; as $\eta \to 0$, the state $\ket{\eta_{1,2}}$ tends to a single photon in pulse mode $\xi_1(\omega)$ according to \eqref{eq:eta_state_fidelity}. Note also that according to \eqref{eq:G_tilde}, the inverse Fourier transform of $\widetilde{G}(\omega)$ must vanish for $t < 0$. In light of the Paley-Wiener criterion \cite{paley1934}, it is impossible for a function and its Fourier transform both to be supported for positive arguments only, meaning that $\eta = 0$ (a strictly localized single photon) is impossible. 

\begin{figure}[tb]
    \includegraphics{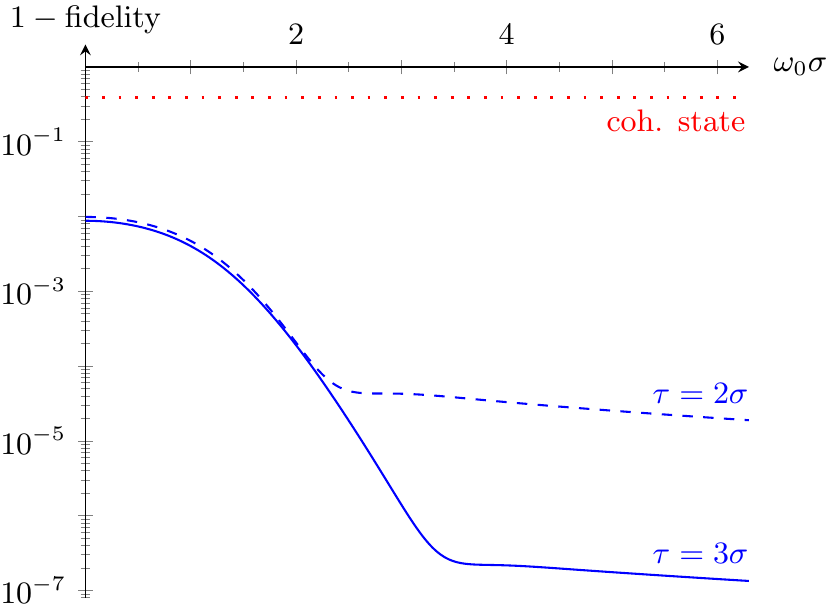}
    \caption{\label{fig:fidelityF}$1 - F$ for the state $\ket{\eta_{1,2}}$ as a function of pulse duration calculated with \eqref{eq:eta_state_fidelity} for a truncated Gaussian seed function \eqref{eq:g_seed_def}: $g(t) \propto u(t) e^{- (t - \tau)^2 / 2 \sigma^2} e^{- i \omega_0 t}$. The pulse delays are $\tau = 2 \sigma$ (blue dashed line) and $\tau = 3 \sigma$ (blue solid line). The red dotted line shows $1 - \text{fidelity} = 1 - 1 / \sqrt{e}$ between a coherent state with parameter $\alpha = 1$ and its single-photon part. Note that in the regime $\omega_0 \sigma \gtrsim 1$, the carrier frequency of the single-photon pulse $\ket{1_1 \kbuff 0_2}$ is approximately equal to $\omega_0$; however, for $\omega_0 \sigma \lesssim 1$, this interpretation is not valid. }
\end{figure}

As an example, let $g(t)$ be a truncated Gaussian with carrier frequency $\omega_0$, duration $\sigma$, and delay $\tau$:
\begin{equation}\label{eq:g_seed_def}
    g(t) \propto u(t) e^{- (t - \tau)^2 / 2\sigma^2} e^{- i\omega_0 t}.
\end{equation}
Here, $u(t)$ is the Heaviside function but with a linearly increasing onset in the interval $0 \leq t \leq 0.1\sigma$ in order to make it continuous. With a numerical Fourier-transform routine, we perform the steps of the algorithm for constructing the state $\ket{\eta_{1,2}}$ to determine $\widetilde{G}(\omega)$ and $\eta$ for various values of the seed-function parameters. Using \eqref{eq:eta_state_fidelity}, we can then calculate the single-photon fidelity $F$ of the state, which is plotted against the seed parameters in \fref{fig:fidelityF}. We see that the fidelity increases rapidly with the pulse duration $\sigma$ but saturates depending on the pulse delay $\tau$. The reason is that a longer pulse contains less negative frequencies, giving a smaller $\eta$, until a point where the truncation dominates the negative-frequency content. For a sufficiently long Gaussian pulse with a sufficiently large delay, $\eta$ can be arbitrarily close to 0, and the state is then near a single photon. 

\section{Expected energy}
The strictly localized state $\ket{\eta_{1,2}}$ is defined on the pulse modes $\xi_1(\omega)$ and $\xi_2(\omega)$, with associated functions $E_1(t)$ and $E_2(t)$ from \eqref{eq:E_n_def}. Define
\begin{subequations}
\begin{align}
    f_0(t)  &= \frac{C}{\sqrt{C^2 - 1}} \lp(E_1^*(t) - \frac{E_2(t)}{C}\rp), \\
    f_1(t)  &= \frac{C}{\sqrt{C^2 - 1}} \lp(E_1(t) + \frac{E_2^*(t)}{C}\rp), \label{eq:f1} \\
    f_2(t)  &= \frac{C}{\sqrt{C^2 - 1}} \lp(E_2(t) + \frac{E_1^*(t)}{C}\rp),
\end{align}
\end{subequations}
where $C^2 = (1 - \eta) / \eta$. Since $g(t)$ vanishes for negative times, it can be shown \footnote{From \eqref{eq:E_n_def}, \eqref{eq:xi_def}, and \eqref{eq:f1}, we have that the Fourier transform of $f_1(t)$ is $f_1(\omega) \propto \mathcal{E}(\omega) \protect\widetilde{G}(\omega)$. The modification in step \ref{itm:G_tilde} of the algorithm for $\protect\ket{\eta_{1,2}}$ has the property that the inverse Fourier transform of $\protect\widetilde{G}(\omega)$ vanishes for negative times since $g(t)$ does. Therefore, $\protect\widetilde{G}(\omega)$ is analytic in the upper half-plane. The function $\mathcal{E}(\omega) = K \sqrt{- i \omega}$ is also analytic here since we can take the branch cut elsewhere. Provided $\protect\widetilde{G}(\omega)$ tends sufficiently fast to zero as $\omega \rightarrow \infty$ in the upper half-plane, we get that $f_1(t) = 0$ for $t < 0$.} that $f_1(t) = 0$ for $t < 0$ as well. The other two functions do not have the same property and can be nonzero for any $t$. 

The operator $S$ from \eqref{eq:S_def} is the two-mode squeeze operator \cite{schumaker1985}, with properties
\begin{align}
    S \ket{0} &= \frac{1}{\cosh\gamma} \sum_{n=0}^\infty (- \tanh\gamma)^n \ket{n_1 \kbuff n_2}, \label{eq:S_vacuum} \\
    S a_1 S^\dagger &= a_1 \cosh \gamma + a_2^\dagger \sinh \gamma, \label{eq:S_a1}
\end{align}
and similar for $a_2$. By \eqref{eq:S_vacuum}, \eqref{eq:S_a1}, and some tedious algebra, we can calculate the expectation value of the normal-ordered product
\begin{equation}\label{eq:expval_norder_2_eta}
\begin{aligned}
    &\qexv{\eta_{1,2}}{\norder{E(t_1)E(t_2)}} \\
    &= \Re \Bigg\{ \frac{4C^2}{C^2 - 1} f_1(t_1) f_1^*(t_2) - f_1(t_1) \Big(f_0(t_2) + 2M f_2(t_2)\Big) \\
    &\qquad\qquad - f_1(t_2) \Big(f_0(t_1) + 2M f_2(t_1)\Big) \Bigg\},
\end{aligned}
\end{equation}
where we have defined the constant
\begin{equation}
    M = \frac{C^2 - 1}{C^2} \sum_{n=0}^\infty C^{-2n + 1} \sqrt{n(n + 1)}.
\end{equation}
Since every term contains a factor $f_1(t)$, it is easily seen that the expectation value \eqref{eq:expval_norder_2_eta} vanishes for negative times. 

\begin{figure}[tb]
    \includegraphics{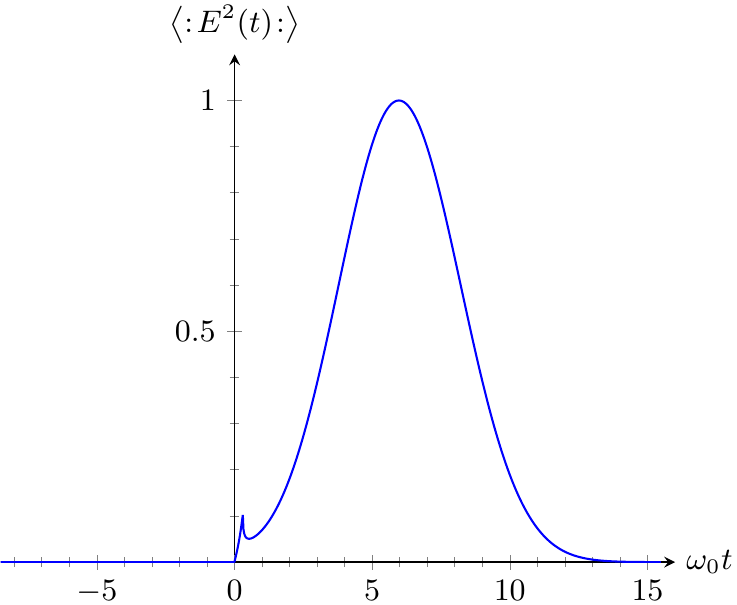}
    \caption{\label{fig:norderE2}Expected normal-ordered energy density $\exv{\norder{E^2(t)}}$ (arb. units) as a function of time for the state $\ket{\eta_{1,2}}$, calculated with \eqref{eq:expval_norder_2_eta}. We have used a truncated Gaussian seed function \eqref{eq:g_seed_def}, as in \fref{fig:fidelityF}, with duration $\omega_0 \sigma = 3$ and delay $\tau = 2 \sigma$. Note that there is no reason to expect this curve to have the same form as the seed function $g(t)$, but it is clearly 0 for $t < 0$ as required by \eqref{expval_vacuum}.}
\end{figure}

By setting $t_1 = t_2$ in \eqref{eq:expval_norder_2_eta}, we find the expected normal-ordered energy density of the strictly localized state $\ket{\eta_{1,2}}$ as a function of time. In \fref{fig:norderE2}, we plot this expectation value for the truncated Gaussian seed function \eqref{eq:g_seed_def} with the same method as in \fref{fig:fidelityF}. The plots clearly indicate that the state is strictly localized to $t \geq 0$ while having a high fidelity with a single photon.

\section{Propagation through barrier/filter}
We now insert a barrier or an optical filter, such as a double prism arrangement \cite{bose1898,keller1999,winful2006}, prism coupler \cite{saleh2007}, waveguide coupler \cite{saleh2007}, waveguide below cutoff \cite{winful2006}, Fabry-Perot interferometer \cite{born2005,saleh2007}, photonic bandgap structure \cite{saleh2007, winful2006}, fiber Bragg grating \cite{saleh2007}, etc. We let the barrier or filter be described as an arbitrary linear system with two interacting modes. The argument can clearly be generalized to any number of modes. 

\begin{figure}[tb]
    \includegraphics{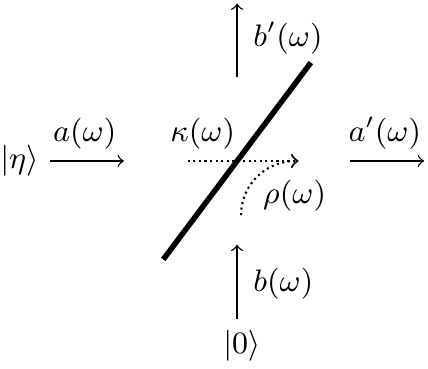}
    \caption{\label{fig:beamsplitter}The optical barrier or filter is modeled as a frequency-dependent, lossless beamsplitter, with signal mode input and output $a(\omega)$ and $a'(\omega)$ respectively, and auxiliary mode input and output $b(\omega)$ and $b'(\omega)$. The reflection coefficient from the right is $\rho(\omega)$, and the transmission coefficient from the left is $\kappa(\omega)$. The auxiliary input is in the vacuum state, and the signal input is some strictly localized state $\ket{\eta}$ incident on the barrier/filter.}
\end{figure}

Linear two-mode interaction can be modeled with a frequency-dependent, lossless beamsplitter, with vacuum incident at the auxiliary input (see \fref{fig:beamsplitter}). We let $a(\omega)$ and $a'(\omega)$ denote the input and output in the signal mode, while $b(\omega)$ and $b'(\omega)$ denote the input and output in the auxiliary mode. A beamsplitter is conveniently described in the Heisenberg picture:
\begin{equation}\label{eq:bs_in_out}
    a'(\omega) = \kappa(\omega) a(\omega) + \rho(\omega) b(\omega),
\end{equation}
for $\omega \geq 0$. Here, $\rho(\omega)$ and $\kappa(\omega)$ are the classical reflection and transmission coefficients, which are defined for all $\omega$ and satisfy $\abs*{\rho(\omega)}^2 + \abs*{\kappa(\omega)}^2 = 1$. 

Due to causality of the beamsplitter, the inverse Fourier transform of the reflection coefficient, $\rho(t)$, and of the transmission coefficient, $\kappa(t)$, must vanish for $t < 0$. Here we have chosen the time reference such that an input signal incident at $t = 0$ will reach the output at $t = 0$; i.e., the reference planes for the input and output modes coincide. The reflection and transmission coefficients also satisfy complex-conjugated symmetry $\rho^*(-\omega) = \rho(\omega)$ and $\kappa^*(-\omega) = \kappa(\omega)$ because the inverse Fourier transforms are real. 

To make use of the beamsplitter relation \eqref{eq:bs_in_out}, we formulate the problem in the Heisenberg picture, with reference \emph{after} the source has produced some strictly localized state $\ket{\eta}$. The electric fields in the input modes are given by \eqref{eq:E_pos}, and we label them $E_a(t)$ and $E_b(t)$ for the signal and auxiliary modes respectively. We then let $E_{a'}(t)$ and $E_{b'}(t)$ be the Heisenberg fields at the output, given by $E_{a'}(t) = U^\dagger E_a(t) U$ and similar for $E_{b'}(t)$. Here, $U$ is the time-evolution operator of the beamsplitter, such that $a'(\omega) = U^\dagger a(\omega) U$ in \eqref{eq:bs_in_out}.

We can decompose the output fields into positive and negative frequencies in the same way as the input fields \eqref{eq:E_pos}:
\begin{align}
    E_{a'}(t)   &= E_{a'}^+(t) + E_{a'}^-(t), \\
    E_{a'}^+(t) &= \int_0^\infty \dd\omega \mathcal{E}(\omega) a'(\omega) e^{-i \omega t},
\end{align}
and similar for $E_{b'}(t)$. From \eqref{eq:bs_in_out}, we also see that $a'(\omega)$ contains only annihilation operators, meaning that normal ordering of $E_{a'}(t)$ and $E_{b'}(t)$ is calculated as usual; e.g.,
\begin{equation}
    \norder{E_{a'}^2(t)}\nordbuffeq = {E_{a'}^+(t)}^2 + {E_{a'}^-(t)}^2 + 2 E_{a'}^-(t) E_{a'}^+(t).
\end{equation}

The local observable we are interested in is the normal-ordered energy density in the signal mode after the action of the beamsplitter $U$. That is, we want to calculate the expectation value $\exv{U^\dagger \norderop{E_a^2(t)} U}$. We can connect this to the Heisenberg fields by noting that the beamsplitter has the property that it leaves the field commutator unchanged: $\comm{E_{a'}^+(t)}{E_{a'}^-(t)} = \comm{E_a^+(t)}{E_a^-(t)}$. This means that the above expectation value can be found directly by the corresponding normal-ordered Heisenberg-picture observable,
\begin{equation}\label{eq:normal_E2}
\begin{aligned}
    &\exv{U^\dagger \norderop{E_a^2(t)} U} \\
    &\quad = \exv{U^\dagger E_a^2(t) U} - \comm{E_a^+(t)}{E_a^-(t)} \\
    &\quad = \exv{E_{a'}^2(t)} - \comm{E_{a'}^+(t)}{E_{a'}^-(t)} \\
    &\quad = \exv{\norder{E_{a'}^2(t)}}.
\end{aligned}
\end{equation}

The source produces a strictly localized state $\ket{\eta}$ incident on the barrier/filter, which corresponds to the signal input of the beamsplitter. The auxiliary input is in the vacuum state $\ket{0}$. It is useful to define operators
\begin{subequations}\label{eq:bs_c_d}
\begin{align}
    c(\omega) &=
    \begin{cases}
        a(\omega),          & \omega > 0, \\
        a^\dagger(-\omega), & \omega < 0,
    \end{cases}\\
    d(\omega) &=
    \begin{cases}
        b(\omega),          & \omega > 0, \\
        b^\dagger(-\omega), & \omega < 0.
    \end{cases}
\end{align}
\end{subequations}
The operators satisfy $c^\dagger(-\omega)=c(\omega)$ and $d^\dagger(-\omega)=d(\omega)$, and obey the same input-output relations \eqref{eq:bs_in_out} as $a(\omega)$ and $b(\omega)$. To see this, define $c'(\omega)$ and $d'(\omega)$ as the corresponding output modes, and we get from \eqref{eq:bs_in_out} and \eqref{eq:bs_c_d} that
\begin{equation}
    c'(\omega) = \kappa(\omega) c(\omega) + \rho(\omega) d(\omega),
\end{equation}
which is now valid for all $\omega$. We can then write the electric field operators at the input as
\begin{subequations}
\begin{align}
    E_a(t) &= \int_{-\infty}^\infty \dd\omega \mathcal{E}(\omega) c(\omega) e^{-i\omega t}, \\
    E_b(t) &= \int_{-\infty}^\infty \dd\omega \mathcal{E}(\omega) d(\omega) e^{-i\omega t},
\end{align}
\end{subequations}
and at the signal output as
\begin{equation}
\begin{aligned}
    E_{a'}(t)   &= \int_{-\infty}^\infty \dd\omega \mathcal{E}(\omega) c'(\omega) e^{-i\omega t} \\
                &= \int_{-\infty}^\infty \dd\omega \mathcal{E}(\omega) \kappa(\omega) c(\omega) e^{-i\omega t} \\
                &+ \int_{-\infty}^\infty \dd\omega \mathcal{E}(\omega) \rho(\omega) d(\omega) e^{-i\omega t} \\
                &= \lp(E_a * \kappa\rp)(t) + \lp(E_b * \rho\rp)(t). \label{eq:Eapconv}
\end{aligned}
\end{equation}

Recall that $\rho(t) = 0$ and $\kappa(t) = 0$ for $t < 0$, so the convolutions in \eqref{eq:Eapconv} mean that $E_a'(t)$ is a superposition of the input fields $E_a(\tau)$ and $E_b(\tau)$ for times $\tau \leq t$. For the state vector $\ket{\eta}\ket{0}$, we therefore see that any local observable will give the vacuum result for $t < 0$. In other words, physical filters preserve causality for any input state strictly localized to $t \geq 0$, including the $\ket{\eta_{1,2}}$ states that are close to single photons. For example, for the $\ket{\eta_{1,2}}$ state, the expected normal-ordered energy density \eqref{eq:normal_E2} becomes
\begin{equation}\label{eq:out_energydens}
\begin{aligned}
    &\qexv{\eta_{1,2} \kbuff 0}{\norder{E_{a'}^2(t)}} \\
    &\quad = \int_0^\infty \dd\tau_1 \kappa(\tau_1) \int_0^\infty \dd\tau_2 \kappa(\tau_2) \\
    &\quad\eqbuff \cdot \qexv{\eta_{1,2}}{\norder{E_a(t - \tau_1) E_a (t - \tau_2)}},
\end{aligned}
\end{equation}
since the auxiliary mode is in the vacuum state so the terms with $E_b(t)$ vanish. Eq.~\eqref{eq:out_energydens} is clearly 0 for $t < 0$ because of the strictly localized property of $\ket{\eta_{1,2}}$.

\section{Numerical examples}

\subsection{Fabry-Perot interferometer}
\begin{figure}[tb]
    \includegraphics{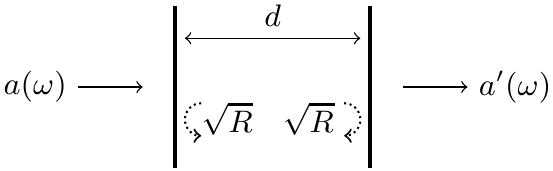}
    \caption{\label{fig:FPint}A Fabry-Perot interferometer consisting of two reflectors, both with power reflection coefficients $R = 0.9$, separated by a distance $d$. There is no phase shift in the reflection off the interior of the reflectors.}
\end{figure}

We would now like to evaluate the full time dependence of \eqref{eq:out_energydens} for some specific examples of optical filters or barriers. Our first example is a Fabry-Perot interferometer consisting of two identical, lossless reflectors with equal power reflection coefficients $R$ and with a spacing $d$ between them (see \fref{fig:FPint}). We assume that there is no phase shift in the reflection as seen from the interior of the interferometer. We also choose the reference planes of the input and output modes to coincide, so that there would be no time delay if the reflectors were absent. By summing multiple reflections, one obtains the following coefficient for transmission through the interferometer:
\begin{equation}\label{eq:FPseries}
\begin{aligned}
    \kappa(\omega)  &= (1 - R)(1 + R e^{i\varphi(\omega)} + R^2 e^{2i\varphi(\omega)} + \dotsb) \\
                    &= \frac{1 - R}{1 - R e^{i\varphi(\omega)}},
\end{aligned}
\end{equation}
where $\varphi(\omega) = 2\omega d/c$ is the round-trip phase, assuming vacuum between the reflectors. Writing $\kappa(\omega) = \abs*{\kappa} e^{\varphi_\kappa}$, the group delay is
\begin{equation}
    \tau_{\text{g}} = \frac{\dd\varphi_\kappa}{\dd\omega} = \frac{2 R (\cos\varphi(\omega) - R)}{R^2 - 2R\cos\varphi(\omega) + 1} \cdot \frac{d}{c}.
\end{equation}
Transforming the reference planes to the physical separation $d$, we get that the group delay in this case is $\tau_{\text{g}} + d/c$. If we interpret this group delay as the distance $d$ divided by a group velocity, we get a group velocity given by
\begin{equation}\label{eq:group_velocity}
    v_{\text{g}} = \frac{R^2 - 2R\cos\varphi(\omega) + 1}{1 - R^2} \cdot c.
\end{equation}
We see that $v_{\text{g}}$ is superluminal when $\tau_{\text{g}} < 0$, which happens off resonance for $\cos\varphi(\omega) < R$, e.g., for frequencies with $\varphi(\omega)$ close to $\pi$. 

Similar to in \fref{fig:norderE2}, we use a numerical routine to calculate the expected normal-ordered energy density before and after the interferometer for the state $\ket{\eta_{1,2}}$ with the truncated Gaussian seed function \eqref{eq:g_seed_def}. We also use that $\kappa(t)$ is in this case a delta-function train (see the series in \eqref{eq:FPseries}) to rewrite the interferometer output expectation value \eqref{eq:out_energydens} as
\begin{equation}\label{eq:out_energydensdiscrete}
\begin{aligned}
    \exv{\norder{E_{a'}^2(t)}}  &= (1 - R)^2 \sum_{n=0}^\infty \sum_{m=0}^\infty R^{n + m} \\
                                & \cdot \qexv{\eta_{1,2}}{\norder{E_a(t - 2nd/c) E_a(t - 2md/c)}}.
\end{aligned}
\end{equation}
In the numerical example, we further choose $R = 0.9$ and $d$ such that $\varphi(\omega_0) = \pi$, which means that a pulse with a sufficiently narrow bandwidth around $\omega_0$ is mostly reflected by the interferometer. For the little light that is transmitted, we expect a superluminal group velocity since $\tau_{\text{g}}$ is negative around $\omega_0$. 

\begin{figure}[tb]
    \includegraphics{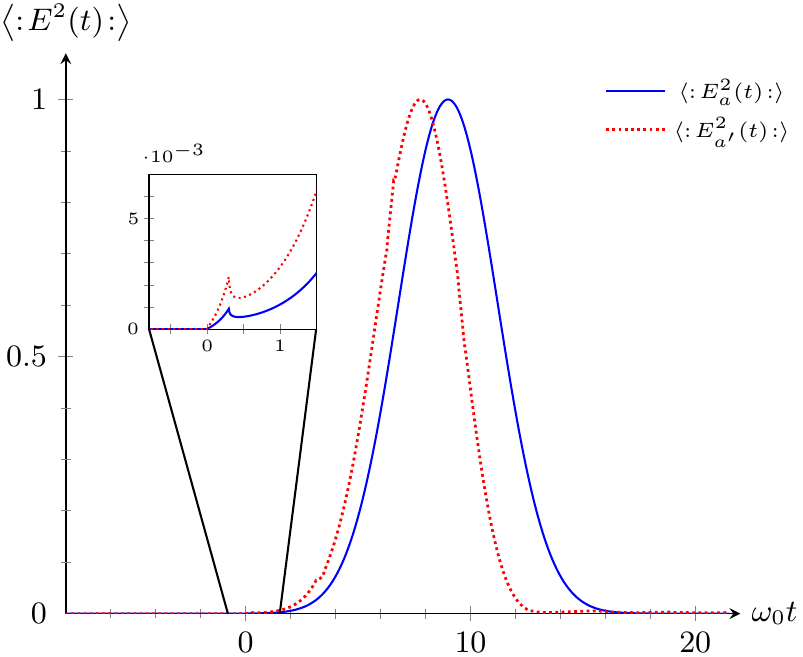}
    \caption{\label{fig:filtered_pulse}Expected normal-ordered energy density (arb. units) as a function of time, before (blue solid line) and after (red dotted line) the interferometer. The state is $\ket{\eta_{1,2}}$ with a truncated Gaussian seed function \eqref{eq:g_seed_def} with $\omega_0 \sigma = 3$ and $\tau = 3 \sigma$. The expectation value is given at the input by \eqref{eq:expval_norder_2_eta} and at the output by \eqref{eq:out_energydensdiscrete}. Both pulse forms are normalized to have a maximum value of 1; the expected value at the output is actually much smaller than that at the input (since most of the light is reflected). The reference planes of the input and output modes coincide, meaning that the expectation value at the output at $t = 0$ corresponds to an actual delay of time $d / c$, implying a causal propagation of the signal.}
\end{figure}

The input and output expectation values are plotted in \fref{fig:filtered_pulse}. Since the input/output reference planes coincide, the plotted output at $t = 0$ corresponds to the output signal emitted from the interferometer at time $d / c$. The figure clearly shows that the output pulse form is shifted to the left of the input pulse form, meaning that the main weight of the pulse arrives at the output before a time $d / c$ has passed, which we interpret as a superluminal group velocity. At the same time, the output plot still starts at $t = 0$, so the leading edge of the signal arrives at the output after time $d / c$, demonstrating that the signal transfer is strictly causal. This is of course similar to the propagation of a classical pulse through the interferometer, but the state here is extremely close to a single photon, as evident from \fref{fig:fidelityF}. 

We can also compare how the expected energy density behaves differently as the incident state is changed. In \fref{fig:comparison}, which was presented in the beginning, we have used the numerical routine to plot the expected normal-ordered energy density at the output of the interferometer for 3 different input states. The interferometer parameters and reference planes are as in \fref{fig:filtered_pulse}, and all the plotted expressions are calculated by the same method as \eqref{eq:out_energydensdiscrete}. The dotted line is the expected energy density at the output for the state $\ket{\eta_{1,2}}$ with a truncated Gaussian seed function \eqref{eq:g_seed_def} with $\omega_0 \sigma = 2.1$ and $\tau = 2.6 \sigma$. The curve clearly shows the same causal signal propagation as in \fref{fig:filtered_pulse}.

The solid line in \fref{fig:comparison} is the expected energy density for a single-photon state with a positive-frequency spectrum given by identifying $\xi_1(\omega)$ from the state $\ket{\eta_{1,2}}$. The single-photon curve matches the curve for $\ket{\eta_{1,2}}$ for large $t > 0$, but as expected, it exhibits a small tail for $t < 0$. We therefore conclude that it cannot be a good representation of the state produced by an on-demand source. Finally, the dashed line in \fref{fig:comparison} represents an approximation where we extend the spectrum of the single-photon pulse to negative frequencies as well. In this approximation, we can avoid the tail for negative $t$ by taking the inverse Fourier transform of the single-photon spectrum, truncating for $t < 0$, and Fourier transforming back again. The plotted curve shows that this method indeed avoids the acausal tail, but the approximation clearly adds significant distortion for large $t > 0$.

\subsection{Photonic bandgap structure}

\begin{figure}[tb]
    \includegraphics{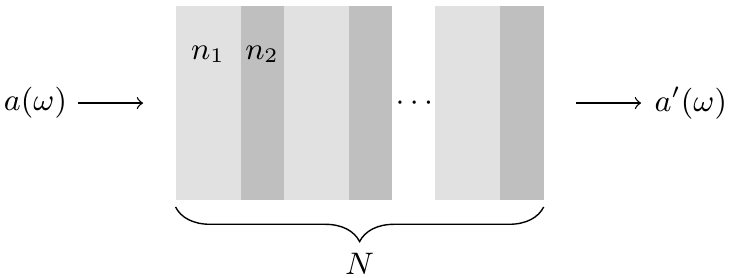}
    \caption{\label{fig:PBGstruct}The photonic bandgap structure consists of $N = 10$ layers with refractive indices $n_1 = 1$ and $n_2 = 2$ (5 pairs of $n_1$-$n_2$ layers). All layers have quarter-wave optical thicknesses for the carrier frequency $\omega_0$.}
\end{figure}

\begin{figure}[tb]
    \includegraphics{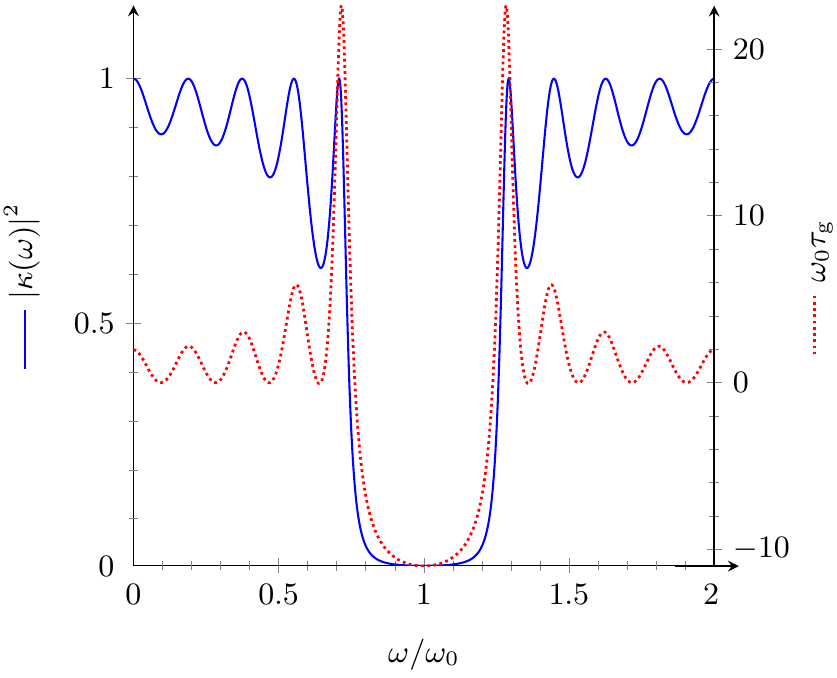}
    \caption{\label{fig:PBGspectrum}The power transmission and group delay spectrum of the photonic bandgap structure. The group delay has been subtracted the delay associated with pure, forward propagation through the structure. The plots show that for frequencies near Bragg reflection resonance, the little light that is transmitted has a negative group delay.}
\end{figure}

\begin{figure}[tb]
    \includegraphics{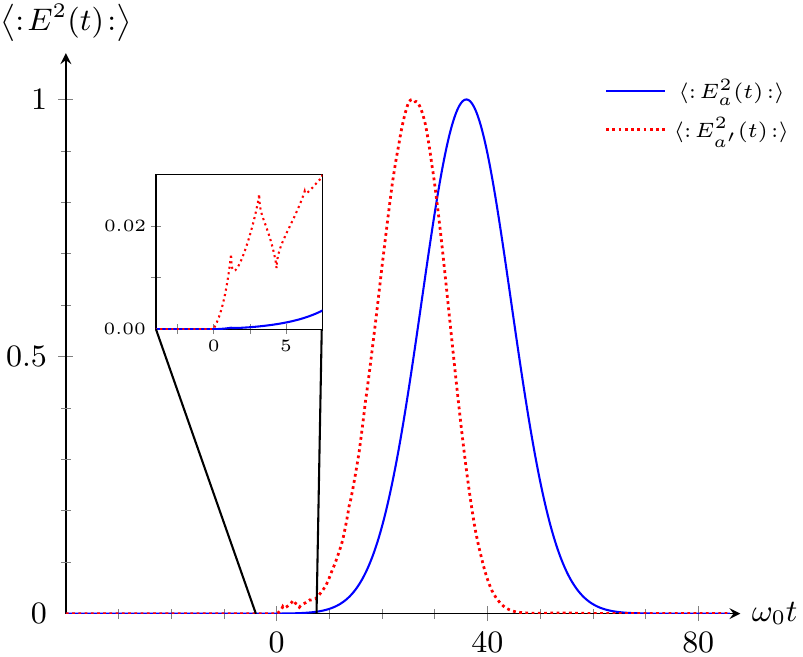}
    \caption{\label{fig:PBGfiltered_pulse}Same as \fref{fig:filtered_pulse} but for the photonic bandgap structure and with $\omega_0\sigma = 12$ and $\tau = 3\sigma$. The expectation value at the output is calculated directly with \eqref{eq:out_energydens}.}
\end{figure}

We now consider a photonic bandgap structure consisting of $N$ layers with refractive indices $n_1$ and $n_2$ (see \fref{fig:PBGstruct}). All layers have quarter-wave thicknesses for the carrier frequency $\omega_0$. The power transmission and group delay spectrums (see \fref{fig:PBGspectrum}) are calculated with explicit expressions involving Chebyshev polynomials (see \cite{born2005}, Ch.~1.6). The input/output reference planes are chosen to coincide, so the group delay is subtracted the delay $N\pi/(2\omega_0)$ of pure, forward propagation through the structure. For frequencies close to Bragg reflection resonance, very little light is transmitted and the group delay is negative, meaning that the group velocity is superluminal. 

As in \fref{fig:filtered_pulse}, we use a numerical routine to find the expected normal-ordered energy density before and after the bandgap structure. The expression for the expectation value at the output \eqref{eq:out_energydens} is simplified to a discrete sum by noting that the bandgap structure has a response that is periodic in frequency with a period of $2 \omega_0$. In the calculations, we set $N = 10$, $n_1 = 1$, and $n_2 = 2$. A nondispersive refractive index $n_2$ is, strictly speaking, not compatible with causality in general. Nevertheless, one may obtain a constant value $> 1$ to any precision with a Lorentzian medium with resonance much larger than the observation frequencies. 

The expectation values are plotted in \fref{fig:PBGfiltered_pulse} and show a similar behavior as in \fref{fig:filtered_pulse}: The main weight of the output pulse is shifted left relative the input pulse, implying a superluminal group velocity. At the same time, as expected from \eqref{eq:out_energydens}, the behavior at the leading edge of the pulse is strictly causal.

\section{Conclusion}
For classical waves, it is well known that information transfer happens at nonanalytic signal points that propagate causally through the system. Single photons, however, have infinite tails and therefore no leading edge, meaning that the question of causality for single photons is problematic. Nevertheless, we can analyze information transfer in the quantum regime by using an on-demand photon source, which produces so-called strictly localized states. Since these states can be very close to single photons while having a leading edge, we can use them for treating questions about causality in optical systems. We demonstrate analytically and numerically that the propagation of single photons in tunneling processes is strictly causal while allowing for superluminal group velocities.



\begin{thebibliography}{34}%
\makeatletter
\providecommand \@ifxundefined [1]{%
 \@ifx{#1\undefined}
}%
\providecommand \@ifnum [1]{%
 \ifnum #1\expandafter \@firstoftwo
 \else \expandafter \@secondoftwo
 \fi
}%
\providecommand \@ifx [1]{%
 \ifx #1\expandafter \@firstoftwo
 \else \expandafter \@secondoftwo
 \fi
}%
\providecommand \natexlab [1]{#1}%
\providecommand \enquote  [1]{``#1''}%
\providecommand \bibnamefont  [1]{#1}%
\providecommand \bibfnamefont [1]{#1}%
\providecommand \citenamefont [1]{#1}%
\providecommand \href@noop [0]{\@secondoftwo}%
\providecommand \href [0]{\begingroup \@sanitize@url \@href}%
\providecommand \@href[1]{\@@startlink{#1}\@@href}%
\providecommand \@@href[1]{\endgroup#1\@@endlink}%
\providecommand \@sanitize@url [0]{\catcode `\\12\catcode `\$12\catcode
  `\&12\catcode `\#12\catcode `\^12\catcode `\_12\catcode `\%12\relax}%
\providecommand \@@startlink[1]{}%
\providecommand \@@endlink[0]{}%
\providecommand \url  [0]{\begingroup\@sanitize@url \@url }%
\providecommand \@url [1]{\endgroup\@href {#1}{\urlprefix }}%
\providecommand \urlprefix  [0]{URL }%
\providecommand \Eprint [0]{\href }%
\providecommand \doibase [0]{https://doi.org/}%
\providecommand \selectlanguage [0]{\@gobble}%
\providecommand \bibinfo  [0]{\@secondoftwo}%
\providecommand \bibfield  [0]{\@secondoftwo}%
\providecommand \translation [1]{[#1]}%
\providecommand \BibitemOpen [0]{}%
\providecommand \bibitemStop [0]{}%
\providecommand \bibitemNoStop [0]{.\EOS\space}%
\providecommand \EOS [0]{\spacefactor3000\relax}%
\providecommand \BibitemShut  [1]{\csname bibitem#1\endcsname}%
\let\auto@bib@innerbib\@empty
\bibitem [{\citenamefont {Hauge}\ and\ \citenamefont
  {St{\o}vneng}(1989)}]{hauge1989}%
  \BibitemOpen
  \bibfield  {author} {\bibinfo {author} {\bibfnamefont {E.~H.}\ \bibnamefont
  {Hauge}}\ and\ \bibinfo {author} {\bibfnamefont {J.~A.}\ \bibnamefont
  {St{\o}vneng}},\ }\bibfield  {title} {\bibinfo {title} {Tunneling times: a
  critical review},\ }\href {https://doi.org/10.1103/RevModPhys.61.917}
  {\bibfield  {journal} {\bibinfo  {journal} {Rev. Mod. Phys.}\ }\textbf
  {\bibinfo {volume} {61}},\ \bibinfo {pages} {917} (\bibinfo {year}
  {1989})}\BibitemShut {NoStop}%
\bibitem [{\citenamefont {Winful}(2006)}]{winful2006}%
  \BibitemOpen
  \bibfield  {author} {\bibinfo {author} {\bibfnamefont {H.~G.}\ \bibnamefont
  {Winful}},\ }\bibfield  {title} {\bibinfo {title} {Tunneling time, the
  {H}artman effect, and superluminality: A proposed resolution of an old
  paradox},\ }\href {https://doi.org/10.1016/j.physrep.2006.09.002} {\bibfield
  {journal} {\bibinfo  {journal} {Phys. Rep.}\ }\textbf {\bibinfo {volume}
  {436}},\ \bibinfo {pages} {1} (\bibinfo {year} {2006})}\BibitemShut {NoStop}%
\bibitem [{\citenamefont {Steinberg}\ \emph {et~al.}(1993)\citenamefont
  {Steinberg}, \citenamefont {Kwiat},\ and\ \citenamefont
  {Chiao}}]{steinberg1993}%
  \BibitemOpen
  \bibfield  {author} {\bibinfo {author} {\bibfnamefont {A.~M.}\ \bibnamefont
  {Steinberg}}, \bibinfo {author} {\bibfnamefont {P.~G.}\ \bibnamefont
  {Kwiat}},\ and\ \bibinfo {author} {\bibfnamefont {R.~Y.}\ \bibnamefont
  {Chiao}},\ }\bibfield  {title} {\bibinfo {title} {Measurement of the
  single-photon tunneling time},\ }\href
  {https://doi.org/10.1103/PhysRevLett.71.708} {\bibfield  {journal} {\bibinfo
  {journal} {Phys. Rev. Lett.}\ }\textbf {\bibinfo {volume} {71}},\ \bibinfo
  {pages} {708} (\bibinfo {year} {1993})}\BibitemShut {NoStop}%
\bibitem [{\citenamefont {Bialynicki-Birula}(1998)}]{bialynicki-birula1998}%
  \BibitemOpen
  \bibfield  {author} {\bibinfo {author} {\bibfnamefont {I.}~\bibnamefont
  {Bialynicki-Birula}},\ }\bibfield  {title} {\bibinfo {title} {Exponential
  localization of photons},\ }\href
  {https://doi.org/10.1103/PhysRevLett.80.5247} {\bibfield  {journal} {\bibinfo
   {journal} {Phys. Rev. Lett.}\ }\textbf {\bibinfo {volume} {80}},\ \bibinfo
  {pages} {5247} (\bibinfo {year} {1998})}\BibitemShut {NoStop}%
\bibitem [{\citenamefont {Wang}\ \emph {et~al.}(2019)\citenamefont {Wang},
  \citenamefont {He}, \citenamefont {Chung}, \citenamefont {Hu}, \citenamefont
  {Yu}, \citenamefont {Chen}, \citenamefont {Ding}, \citenamefont {Chen},
  \citenamefont {Qin}, \citenamefont {Yang}, \citenamefont {Liu}, \citenamefont
  {Duan}, \citenamefont {Li}, \citenamefont {Gerhardt}, \citenamefont
  {Winkler}, \citenamefont {Jurkat}, \citenamefont {Wang}, \citenamefont
  {Gregersen}, \citenamefont {Huo}, \citenamefont {Dai}, \citenamefont {Yu},
  \citenamefont {H{\"o}fling}, \citenamefont {Lu},\ and\ \citenamefont
  {Pan}}]{wang2019}%
  \BibitemOpen
  \bibfield  {author} {\bibinfo {author} {\bibfnamefont {H.}~\bibnamefont
  {Wang}}, \bibinfo {author} {\bibfnamefont {Y.-M.}\ \bibnamefont {He}},
  \bibinfo {author} {\bibfnamefont {T.-H.}\ \bibnamefont {Chung}}, \bibinfo
  {author} {\bibfnamefont {H.}~\bibnamefont {Hu}}, \bibinfo {author}
  {\bibfnamefont {Y.}~\bibnamefont {Yu}}, \bibinfo {author} {\bibfnamefont
  {S.}~\bibnamefont {Chen}}, \bibinfo {author} {\bibfnamefont {X.}~\bibnamefont
  {Ding}}, \bibinfo {author} {\bibfnamefont {M.-C.}\ \bibnamefont {Chen}},
  \bibinfo {author} {\bibfnamefont {J.}~\bibnamefont {Qin}}, \bibinfo {author}
  {\bibfnamefont {X.}~\bibnamefont {Yang}}, \bibinfo {author} {\bibfnamefont
  {R.-Z.}\ \bibnamefont {Liu}}, \bibinfo {author} {\bibfnamefont {Z.-C.}\
  \bibnamefont {Duan}}, \bibinfo {author} {\bibfnamefont {J.-P.}\ \bibnamefont
  {Li}}, \bibinfo {author} {\bibfnamefont {S.}~\bibnamefont {Gerhardt}},
  \bibinfo {author} {\bibfnamefont {K.}~\bibnamefont {Winkler}}, \bibinfo
  {author} {\bibfnamefont {J.}~\bibnamefont {Jurkat}}, \bibinfo {author}
  {\bibfnamefont {L.-J.}\ \bibnamefont {Wang}}, \bibinfo {author}
  {\bibfnamefont {N.}~\bibnamefont {Gregersen}}, \bibinfo {author}
  {\bibfnamefont {Y.-H.}\ \bibnamefont {Huo}}, \bibinfo {author} {\bibfnamefont
  {Q.}~\bibnamefont {Dai}}, \bibinfo {author} {\bibfnamefont {S.}~\bibnamefont
  {Yu}}, \bibinfo {author} {\bibfnamefont {S.}~\bibnamefont {H{\"o}fling}},
  \bibinfo {author} {\bibfnamefont {C.-Y.}\ \bibnamefont {Lu}},\ and\ \bibinfo
  {author} {\bibfnamefont {J.-W.}\ \bibnamefont {Pan}},\ }\bibfield  {title}
  {\bibinfo {title} {Towards optimal single-photon sources from polarized
  microcavities},\ }\href {https://doi.org/10.1038/s41566-019-0494-3}
  {\bibfield  {journal} {\bibinfo  {journal} {Nat. Photon.}\ }\textbf {\bibinfo
  {volume} {13}},\ \bibinfo {pages} {770} (\bibinfo {year} {2019})},\ \Eprint
  {https://arxiv.org/abs/1907.06818} {arXiv:1907.06818 [quant-ph]} \BibitemShut
  {NoStop}%
\bibitem [{\citenamefont {Scheel}(2009)}]{scheel2009}%
  \BibitemOpen
  \bibfield  {author} {\bibinfo {author} {\bibfnamefont {S.}~\bibnamefont
  {Scheel}},\ }\bibfield  {title} {\bibinfo {title} {Single-photon sources --
  an introduction},\ }\href {https://doi.org/10.1080/09500340802331849}
  {\bibfield  {journal} {\bibinfo  {journal} {J. Mod. Opt.}\ }\textbf {\bibinfo
  {volume} {56}},\ \bibinfo {pages} {141} (\bibinfo {year} {2009})}\BibitemShut
  {NoStop}%
\bibitem [{\citenamefont {Eisaman}\ \emph {et~al.}(2011)\citenamefont
  {Eisaman}, \citenamefont {Fan}, \citenamefont {Migdall},\ and\ \citenamefont
  {Polyakov}}]{eisaman2011}%
  \BibitemOpen
  \bibfield  {author} {\bibinfo {author} {\bibfnamefont {M.~D.}\ \bibnamefont
  {Eisaman}}, \bibinfo {author} {\bibfnamefont {J.}~\bibnamefont {Fan}},
  \bibinfo {author} {\bibfnamefont {A.}~\bibnamefont {Migdall}},\ and\ \bibinfo
  {author} {\bibfnamefont {S.~V.}\ \bibnamefont {Polyakov}},\ }\bibfield
  {title} {\bibinfo {title} {Invited review article: Single-photon sources and
  detectors},\ }\href {https://doi.org/10.1063/1.3610677} {\bibfield  {journal}
  {\bibinfo  {journal} {Rev. Sci. Instrum.}\ }\textbf {\bibinfo {volume}
  {82}},\ \bibinfo {pages} {071101} (\bibinfo {year} {2011})}\BibitemShut
  {NoStop}%
\bibitem [{\citenamefont {Senellart}\ \emph {et~al.}(2017)\citenamefont
  {Senellart}, \citenamefont {Solomon},\ and\ \citenamefont
  {White}}]{senellart2017}%
  \BibitemOpen
  \bibfield  {author} {\bibinfo {author} {\bibfnamefont {P.}~\bibnamefont
  {Senellart}}, \bibinfo {author} {\bibfnamefont {G.}~\bibnamefont {Solomon}},\
  and\ \bibinfo {author} {\bibfnamefont {A.}~\bibnamefont {White}},\ }\bibfield
   {title} {\bibinfo {title} {High-performance semiconductor quantum-dot
  single-photon sources},\ }\href {https://doi.org/10.1038/nnano.2017.218}
  {\bibfield  {journal} {\bibinfo  {journal} {Nat. Nanotechnol.}\ }\textbf
  {\bibinfo {volume} {12}},\ \bibinfo {pages} {1026} (\bibinfo {year}
  {2017})}\BibitemShut {NoStop}%
\bibitem [{\citenamefont {Sinha}\ \emph {et~al.}(2019)\citenamefont {Sinha},
  \citenamefont {Sahoo}, \citenamefont {Singh}, \citenamefont {Joarder},
  \citenamefont {Chatterjee},\ and\ \citenamefont {Chakraborti}}]{sinha2019}%
  \BibitemOpen
  \bibfield  {author} {\bibinfo {author} {\bibfnamefont {U.}~\bibnamefont
  {Sinha}}, \bibinfo {author} {\bibfnamefont {S.~N.}\ \bibnamefont {Sahoo}},
  \bibinfo {author} {\bibfnamefont {A.}~\bibnamefont {Singh}}, \bibinfo
  {author} {\bibfnamefont {K.}~\bibnamefont {Joarder}}, \bibinfo {author}
  {\bibfnamefont {R.}~\bibnamefont {Chatterjee}},\ and\ \bibinfo {author}
  {\bibfnamefont {S.}~\bibnamefont {Chakraborti}},\ }\bibfield  {title}
  {\bibinfo {title} {Single-photon sources},\ }\href
  {https://doi.org/10.1364/OPN.30.9.000032} {\bibfield  {journal} {\bibinfo
  {journal} {Opt. Photonics News}\ }\textbf {\bibinfo {volume} {30}},\ \bibinfo
  {pages} {32} (\bibinfo {year} {2019})},\ \Eprint
  {https://arxiv.org/abs/1906.09565} {arXiv:1906.09565 [quant-ph]} \BibitemShut
  {NoStop}%
\bibitem [{\citenamefont {Knight}(1961)}]{knight1961}%
  \BibitemOpen
  \bibfield  {author} {\bibinfo {author} {\bibfnamefont {J.~M.}\ \bibnamefont
  {Knight}},\ }\bibfield  {title} {\bibinfo {title} {Strict localization in
  quantum field theory},\ }\href {https://doi.org/10.1063/1.1703731} {\bibfield
   {journal} {\bibinfo  {journal} {J. Math. Phys.}\ }\textbf {\bibinfo {volume}
  {2}},\ \bibinfo {pages} {459} (\bibinfo {year} {1961})}\BibitemShut {NoStop}%
\bibitem [{\citenamefont {Licht}(1963)}]{licht1963}%
  \BibitemOpen
  \bibfield  {author} {\bibinfo {author} {\bibfnamefont {A.~L.}\ \bibnamefont
  {Licht}},\ }\bibfield  {title} {\bibinfo {title} {Strict localization},\
  }\href {https://doi.org/10.1063/1.1703925} {\bibfield  {journal} {\bibinfo
  {journal} {J. Math. Phys.}\ }\textbf {\bibinfo {volume} {4}},\ \bibinfo
  {pages} {1443} (\bibinfo {year} {1963})}\BibitemShut {NoStop}%
\bibitem [{\citenamefont {Gulla}\ and\ \citenamefont
  {Skaar}(2021)}]{gulla2021}%
  \BibitemOpen
  \bibfield  {author} {\bibinfo {author} {\bibfnamefont {J.}~\bibnamefont
  {Gulla}}\ and\ \bibinfo {author} {\bibfnamefont {J.}~\bibnamefont {Skaar}},\
  }\bibfield  {title} {\bibinfo {title} {Approaching single-photon pulses},\
  }\href {https://doi.org/10.1103/PhysRevLett.126.073601} {\bibfield  {journal}
  {\bibinfo  {journal} {Phys. Rev. Lett.}\ }\textbf {\bibinfo {volume} {126}},\
  \bibinfo {pages} {073601} (\bibinfo {year} {2021})},\ \Eprint
  {https://arxiv.org/abs/2008.07483} {arXiv:2008.07483 [quant-ph]} \BibitemShut
  {NoStop}%
\bibitem [{\citenamefont {Milonni}\ \emph {et~al.}(1995)\citenamefont
  {Milonni}, \citenamefont {James},\ and\ \citenamefont {Fearn}}]{milonni1995}%
  \BibitemOpen
  \bibfield  {author} {\bibinfo {author} {\bibfnamefont {P.~W.}\ \bibnamefont
  {Milonni}}, \bibinfo {author} {\bibfnamefont {D.~F.~V.}\ \bibnamefont
  {James}},\ and\ \bibinfo {author} {\bibfnamefont {H.}~\bibnamefont {Fearn}},\
  }\bibfield  {title} {\bibinfo {title} {Photodetection and causality in
  quantum optics},\ }\href {https://doi.org/10.1103/PhysRevA.52.1525}
  {\bibfield  {journal} {\bibinfo  {journal} {Phys. Rev. A}\ }\textbf {\bibinfo
  {volume} {52}},\ \bibinfo {pages} {1525} (\bibinfo {year}
  {1995})}\BibitemShut {NoStop}%
\bibitem [{\citenamefont {Milonni}(1994)}]{milonni1994}%
  \BibitemOpen
  \bibfield  {author} {\bibinfo {author} {\bibfnamefont {P.~W.}\ \bibnamefont
  {Milonni}},\ }\href {https://doi.org/10.1016/C2009-0-21295-5} {\emph
  {\bibinfo {title} {The Quantum Vacuum: An Introduction to Quantum
  Electrodynamics}}}\ (\bibinfo  {publisher} {Academic Press},\ \bibinfo
  {address} {Boston},\ \bibinfo {year} {1994})\ Chap.\ \bibinfo {chapter}
  {4.6}\BibitemShut {NoStop}%
\bibitem [{\citenamefont {Ferretti}(1968)}]{ferretti1968}%
  \BibitemOpen
  \bibfield  {author} {\bibinfo {author} {\bibfnamefont {B.}~\bibnamefont
  {Ferretti}},\ }\bibfield  {title} {\bibinfo {title} {Propagation of signals
  and particles: A volume dedicated to {G}ilberto {B}ernardini in his sixtieth
  birthday},\ }in\ \href {https://doi.org/10.1016/b978-0-12-395657-6.50011-2}
  {\emph {\bibinfo {booktitle} {Old and New Problems in Elementary
  Particles}}},\ \bibinfo {editor} {edited by\ \bibinfo {editor} {\bibfnamefont
  {G.}~\bibnamefont {Puppi}}}\ (\bibinfo  {publisher} {Academic Press},\
  \bibinfo {address} {New York},\ \bibinfo {year} {1968})\ pp.\ \bibinfo
  {pages} {108--119}\BibitemShut {NoStop}%
\bibitem [{\citenamefont {Keller}(1999)}]{keller1999}%
  \BibitemOpen
  \bibfield  {author} {\bibinfo {author} {\bibfnamefont {O.}~\bibnamefont
  {Keller}},\ }\bibfield  {title} {\bibinfo {title} {Relation between spatial
  confinement of light and optical tunneling},\ }\href
  {https://doi.org/10.1103/physreva.60.1652} {\bibfield  {journal} {\bibinfo
  {journal} {Phys. Rev. A}\ }\textbf {\bibinfo {volume} {60}},\ \bibinfo
  {pages} {1652} (\bibinfo {year} {1999})}\BibitemShut {NoStop}%
\bibitem [{\citenamefont {Saari}\ \emph {et~al.}(2005)\citenamefont {Saari},
  \citenamefont {Menert},\ and\ \citenamefont {Valtna}}]{saari2005}%
  \BibitemOpen
  \bibfield  {author} {\bibinfo {author} {\bibfnamefont {P.}~\bibnamefont
  {Saari}}, \bibinfo {author} {\bibfnamefont {M.}~\bibnamefont {Menert}},\ and\
  \bibinfo {author} {\bibfnamefont {H.}~\bibnamefont {Valtna}},\ }\bibfield
  {title} {\bibinfo {title} {Photon localization barrier can be overcome},\
  }\href {https://doi.org/10.1016/j.optcom.2004.11.020} {\bibfield  {journal}
  {\bibinfo  {journal} {Opt. Commun.}\ }\textbf {\bibinfo {volume} {246}},\
  \bibinfo {pages} {445} (\bibinfo {year} {2005})},\ \Eprint
  {https://arxiv.org/abs/quant-ph/0409034} {arXiv:quant-ph/0409034}
  \BibitemShut {NoStop}%
\bibitem [{Note1()}]{Note1}%
  \BibitemOpen
  \bibinfo {note} {For calculating normal-ordered expectation values, we may
  omit modes in the field expression for which the state contains vacuum.
  Choosing to omit leftward-moving modes is in principle no different than
  omitting the other 3D directions when considering 1D propagation; it is
  anyway just a choice since we want to find an example of a strictly localized
  state for which the other modes are vacuum. There could very well be other
  solutions of states strictly localized to $t \geq 0$ that contain both left-
  and right-moving components (or other 3D modes), but we show that solutions
  exist with purely rightward-moving modes.}\BibitemShut {Stop}%
\bibitem [{\citenamefont {Paley}\ and\ \citenamefont
  {Wiener}(1934)}]{paley1934}%
  \BibitemOpen
  \bibfield  {author} {\bibinfo {author} {\bibfnamefont {R.~E. A.~C.}\
  \bibnamefont {Paley}}\ and\ \bibinfo {author} {\bibfnamefont
  {N.}~\bibnamefont {Wiener}},\ }\href {https://doi.org/10.1090/coll/019}
  {\emph {\bibinfo {title} {{F}ourier Transforms in the Complex Domain}}},\
  {A}merican Mathematical Society Colloquium Publications Vol. 19\ (\bibinfo
  {publisher} {American Mathematical Society},\ \bibinfo {address} {New York},\
  \bibinfo {year} {1934})\ Chap.\ \bibinfo {chapter} {I.7}\BibitemShut
  {NoStop}%
\bibitem [{\citenamefont {Newton}\ and\ \citenamefont
  {Wigner}(1949)}]{newton1949}%
  \BibitemOpen
  \bibfield  {author} {\bibinfo {author} {\bibfnamefont {T.~D.}\ \bibnamefont
  {Newton}}\ and\ \bibinfo {author} {\bibfnamefont {E.~P.}\ \bibnamefont
  {Wigner}},\ }\bibfield  {title} {\bibinfo {title} {Localized states for
  elementary systems},\ }\href {https://doi.org/10.1103/revmodphys.21.400}
  {\bibfield  {journal} {\bibinfo  {journal} {Rev. Mod. Phys.}\ }\textbf
  {\bibinfo {volume} {21}},\ \bibinfo {pages} {400} (\bibinfo {year}
  {1949})}\BibitemShut {NoStop}%
\bibitem [{\citenamefont {Jauch}\ and\ \citenamefont
  {Piron}(1967)}]{jauch1967}%
  \BibitemOpen
  \bibfield  {author} {\bibinfo {author} {\bibfnamefont {J.~M.}\ \bibnamefont
  {Jauch}}\ and\ \bibinfo {author} {\bibfnamefont {C.}~\bibnamefont {Piron}},\
  }\bibfield  {title} {\bibinfo {title} {Generalized localizability},\ }\href
  {https://doi.org/10.5169/SEALS-113783} {\bibfield  {journal} {\bibinfo
  {journal} {{H}elv. Phys. Acta}\ }\textbf {\bibinfo {volume} {40}},\ \bibinfo
  {pages} {559} (\bibinfo {year} {1967})}\BibitemShut {NoStop}%
\bibitem [{\citenamefont {Amrein}(1969)}]{amrein1969}%
  \BibitemOpen
  \bibfield  {author} {\bibinfo {author} {\bibfnamefont {W.~O.}\ \bibnamefont
  {Amrein}},\ }\bibfield  {title} {\bibinfo {title} {Localizability for
  particles of mass zero},\ }\href@noop {} {\bibfield  {journal} {\bibinfo
  {journal} {{H}elv. Phys. Acta}\ }\textbf {\bibinfo {volume} {42}},\ \bibinfo
  {pages} {149} (\bibinfo {year} {1969})}\BibitemShut {NoStop}%
\bibitem [{\citenamefont {Adlard}\ \emph {et~al.}(1997)\citenamefont {Adlard},
  \citenamefont {Pike},\ and\ \citenamefont {Sarkar}}]{adlard1997}%
  \BibitemOpen
  \bibfield  {author} {\bibinfo {author} {\bibfnamefont {C.}~\bibnamefont
  {Adlard}}, \bibinfo {author} {\bibfnamefont {E.~R.}\ \bibnamefont {Pike}},\
  and\ \bibinfo {author} {\bibfnamefont {S.}~\bibnamefont {Sarkar}},\
  }\bibfield  {title} {\bibinfo {title} {Localization of one-photon states},\
  }\href {https://doi.org/10.1103/physrevlett.79.1585} {\bibfield  {journal}
  {\bibinfo  {journal} {Phys. Rev. Lett.}\ }\textbf {\bibinfo {volume} {79}},\
  \bibinfo {pages} {1585} (\bibinfo {year} {1997})}\BibitemShut {NoStop}%
\bibitem [{\citenamefont {Ciattoni}\ and\ \citenamefont
  {Conti}(2007)}]{ciattoni2007}%
  \BibitemOpen
  \bibfield  {author} {\bibinfo {author} {\bibfnamefont {A.}~\bibnamefont
  {Ciattoni}}\ and\ \bibinfo {author} {\bibfnamefont {C.}~\bibnamefont
  {Conti}},\ }\bibfield  {title} {\bibinfo {title} {Quantum electromagnetic {X}
  waves},\ }\href {https://doi.org/10.1364/josab.24.002195} {\bibfield
  {journal} {\bibinfo  {journal} {J. Opt. Soc. {A}m. B}\ }\textbf {\bibinfo
  {volume} {24}},\ \bibinfo {pages} {2195} (\bibinfo {year} {2007})},\ \Eprint
  {https://arxiv.org/abs/0704.0442} {arXiv:0704.0442 [physics.optics]}
  \BibitemShut {NoStop}%
\bibitem [{\citenamefont {Saari}\ and\ \citenamefont
  {Reivelt}(1997)}]{saari1997}%
  \BibitemOpen
  \bibfield  {author} {\bibinfo {author} {\bibfnamefont {P.}~\bibnamefont
  {Saari}}\ and\ \bibinfo {author} {\bibfnamefont {K.}~\bibnamefont
  {Reivelt}},\ }\bibfield  {title} {\bibinfo {title} {Evidence of {X}-shaped
  propagation-invariant localized light waves},\ }\href
  {https://doi.org/10.1103/physrevlett.79.4135} {\bibfield  {journal} {\bibinfo
   {journal} {Phys. Rev. Lett.}\ }\textbf {\bibinfo {volume} {79}},\ \bibinfo
  {pages} {4135} (\bibinfo {year} {1997})}\BibitemShut {NoStop}%
\bibitem [{\citenamefont {Loudon}(2001)}]{loudon2001}%
  \BibitemOpen
  \bibfield  {author} {\bibinfo {author} {\bibfnamefont {R.}~\bibnamefont
  {Loudon}},\ }\href@noop {} {\emph {\bibinfo {title} {The Quantum Theory of
  Light}}},\ \bibinfo {edition} {3rd}\ ed.\ (\bibinfo  {publisher} {Oxford
  University Press},\ \bibinfo {address} {Oxford},\ \bibinfo {year}
  {2001})\BibitemShut {NoStop}%
\bibitem [{\citenamefont {Tatarskii}(1990)}]{tatarskii1990}%
  \BibitemOpen
  \bibfield  {author} {\bibinfo {author} {\bibfnamefont {V.~I.}\ \bibnamefont
  {Tatarskii}},\ }\bibfield  {title} {\bibinfo {title} {Corrections to the
  theory of photocounting},\ }\href
  {https://doi.org/10.1016/0375-9601(90)90521-O} {\bibfield  {journal}
  {\bibinfo  {journal} {Phys. Lett. A}\ }\textbf {\bibinfo {volume} {144}},\
  \bibinfo {pages} {491} (\bibinfo {year} {1990})}\BibitemShut {NoStop}%
\bibitem [{Note2()}]{Note2}%
  \BibitemOpen
  \bibinfo {note} {The inapplicability of the intensity operator for rapidly
  varying envelopes is not so different from the situation in classical
  electrodynamics. For example, writing the electric field as $E(t) = \Re
  {E_0(t) e^{-i\omega _0 t}}$ with carrier $\omega _0$ and envelope $E_0(t)$,
  the electric energy density is $\protect \frac {1}{2} \epsilon _0 E^2(t) =
  \protect \frac {1}{8}\epsilon _0 \mathopen {}\left [E_0(t)^2 e^{-2i\omega _0
  t} + E_0^*(t)^2e^{2i\omega _0 t} + 2E_0^*(t) E_0(t) \right ]$, where
  $\epsilon _0$ is the permittivity. In the quasi-monochromatic case where the
  envelope $E_0(t)$ varies slowly, we can average over an optical period such
  that the first two terms are washed out. This leads to a result proportional
  to $E_0^*(t) E_0(t)$. However, when the envelope varies rapidly, we must
  retain the first two terms, keeping the original expression $\protect \frac
  {1}{2}\epsilon _0 E^2(t)$.}\BibitemShut {Stop}%
\bibitem [{\citenamefont {Su}\ \emph {et~al.}(2016)\citenamefont {Su},
  \citenamefont {Chinnarasu}, \citenamefont {Kuo},\ and\ \citenamefont
  {Chuu}}]{su2016}%
  \BibitemOpen
  \bibfield  {author} {\bibinfo {author} {\bibfnamefont {W.-M.}\ \bibnamefont
  {Su}}, \bibinfo {author} {\bibfnamefont {R.}~\bibnamefont {Chinnarasu}},
  \bibinfo {author} {\bibfnamefont {C.-H.}\ \bibnamefont {Kuo}},\ and\ \bibinfo
  {author} {\bibfnamefont {C.-S.}\ \bibnamefont {Chuu}},\ }\bibfield  {title}
  {\bibinfo {title} {Shaping single photons and biphotons by inherent losses},\
  }\href {https://doi.org/10.1103/PhysRevA.94.033805} {\bibfield  {journal}
  {\bibinfo  {journal} {Phys. Rev. A}\ }\textbf {\bibinfo {volume} {94}},\
  \bibinfo {pages} {033805} (\bibinfo {year} {2016})},\ \Eprint
  {https://arxiv.org/abs/1609.00761} {arXiv:1609.00761 [quant-ph]} \BibitemShut
  {NoStop}%
\bibitem [{Note3()}]{Note3}%
  \BibitemOpen
  \bibinfo {note} {From \protect \textup {\hbox {\mathsurround \z@ \protect
  \normalfont (\ignorespaces \ref {eq:E_n_def}\unskip \@@italiccorr )}},
  \protect \textup {\hbox {\mathsurround \z@ \protect \normalfont
  (\ignorespaces \ref {eq:xi_def}\unskip \@@italiccorr )}}, and \protect
  \textup {\hbox {\mathsurround \z@ \protect \normalfont (\ignorespaces \ref
  {eq:f1}\unskip \@@italiccorr )}}, we have that the Fourier transform of
  $f_1(t)$ is $f_1(\omega ) \propto \protect \mathcal {E}(\omega ) \protect
  \widetilde {G}(\omega )$. The modification in step \ref {itm:G_tilde} of the
  algorithm for $\protect \ket {\eta _{1,2}}$ has the property that the inverse
  Fourier transform of $\protect \widetilde {G}(\omega )$ vanishes for negative
  times since $g(t)$ does. Therefore, $\protect \widetilde {G}(\omega )$ is
  analytic in the upper half-plane. The function $\protect \mathcal {E}(\omega
  ) = K \protect \sqrt {- i \omega }$ is also analytic here since we can take
  the branch cut elsewhere. Provided $\protect \widetilde {G}(\omega )$ tends
  sufficiently fast to zero as $\omega \rightarrow \infty $ in the upper
  half-plane, we get that $f_1(t) = 0$ for $t < 0$.}\BibitemShut {Stop}%
\bibitem [{\citenamefont {Schumaker}\ and\ \citenamefont
  {Caves}(1985)}]{schumaker1985}%
  \BibitemOpen
  \bibfield  {author} {\bibinfo {author} {\bibfnamefont {B.~L.}\ \bibnamefont
  {Schumaker}}\ and\ \bibinfo {author} {\bibfnamefont {C.~M.}\ \bibnamefont
  {Caves}},\ }\bibfield  {title} {\bibinfo {title} {New formalism for
  two-photon quantum optics. {II}. {M}athematical foundation and compact
  notation},\ }\href {https://doi.org/10.1103/physreva.31.3093} {\bibfield
  {journal} {\bibinfo  {journal} {Phys. Rev. A}\ }\textbf {\bibinfo {volume}
  {31}},\ \bibinfo {pages} {3093} (\bibinfo {year} {1985})}\BibitemShut
  {NoStop}%
\bibitem [{\citenamefont {Bose}(1898)}]{bose1898}%
  \BibitemOpen
  \bibfield  {author} {\bibinfo {author} {\bibfnamefont {J.~C.}\ \bibnamefont
  {Bose}},\ }\bibfield  {title} {\bibinfo {title} {On the influence of the
  thickness of air-space on total reflection of electric radiation},\ }\href
  {https://doi.org/10.1098/rspl.1897.0114} {\bibfield  {journal} {\bibinfo
  {journal} {Proc. R. Soc. {L}ondon}\ }\textbf {\bibinfo {volume} {62}},\
  \bibinfo {pages} {300} (\bibinfo {year} {1898})}\BibitemShut {NoStop}%
\bibitem [{\citenamefont {Saleh}\ and\ \citenamefont
  {Teich}(2007)}]{saleh2007}%
  \BibitemOpen
  \bibfield  {author} {\bibinfo {author} {\bibfnamefont {B.~E.~A.}\
  \bibnamefont {Saleh}}\ and\ \bibinfo {author} {\bibfnamefont {M.~C.}\
  \bibnamefont {Teich}},\ }\href@noop {} {\emph {\bibinfo {title} {Fundamentals
  of Photonics}}},\ \bibinfo {edition} {2nd}\ ed.,\ Wiley Series in Pure and
  Applied Optics\ (\bibinfo  {publisher} {Wiley-Interscience},\ \bibinfo
  {address} {Hoboken},\ \bibinfo {year} {2007})\BibitemShut {NoStop}%
\bibitem [{\citenamefont {Born}\ and\ \citenamefont {Wolf}(2005)}]{born2005}%
  \BibitemOpen
  \bibfield  {author} {\bibinfo {author} {\bibfnamefont {M.}~\bibnamefont
  {Born}}\ and\ \bibinfo {author} {\bibfnamefont {E.}~\bibnamefont {Wolf}},\
  }\href {https://doi.org/10.1017/9781108769914} {\emph {\bibinfo {title}
  {Principles of Optics: Electromagnetic Theory of Propagation, Interference
  and Diffraction of Light}}},\ \bibinfo {edition} {7th}\ ed.\ (\bibinfo
  {publisher} {Cambridge University Press},\ \bibinfo {address} {Cambridge},\
  \bibinfo {year} {2005})\BibitemShut {NoStop}%
\end{thebibliography}

%

\onecolumngrid
\end{document}